# Anomalous Hall effect of light-driven three-dimensional Dirac electrons in bismuth

Yoshua Hirai[1], Naotaka Yoshikawa[1], Masashi Kawaguchi[1], Masamitsu Hayashi[1], Shun Okumura[2], Takashi Oka[2], and Ryo Shimano[1,3]

[1]*Department of Physics, The University of Tokyo, Hongo, Tokyo, 113-0033, Japan*
[2]*Insitute for Solid State Physics, The University of Tokyo, Kashiwa, Chiba, 277-8581, Japan*
[3]*Cryogenic Research Center, The University of Tokyo, Yayoi, Tokyo 113-0032, Japan*

**Abstract**

Recent advancement in laser technology has opened the path toward the manipulation of functionalities in quantum materials by intense coherent light. Here, we study three-dimensional (3D) Dirac electrons driven by circularly polarized light (CPL), when the photon energy lies within the Dirac bands. As an experimental realization of this setup, we irradiate a thin film sample of elemental bismuth, which is a well-known semimetal hosting 3D Dirac electrons, with mid-infrared CPL. We successfully observe the emergence of the anomalous Hall effect (AHE) via terahertz Faraday rotation that is both pump-helicity-dependent and instantaneous. We compare our experimental findings with the results of Floquet theory, which is a powerful framework for analyzing the electronic band structure driven by coherent light. The contribution from the band structures near the one-photon resonant positions to the AHE shows a field-strength dependence consistent with our experimental results. The effective Hamiltonian on which we base our model calculations also implies that a pair of "double Weyl points" emerge due to the CPL-induced hybridization between the occupied and unoccupied 3D Dirac bands. Our findings shed light on ultrafast control of material properties in nonlinear topological optics.

**Main text**

## I. Introduction

Optical manipulation of quantum materials by intense coherent laser light has gained growing interest as a unique tool to realize new quantum phases with no equilibrium counterpart. Of various protocols, Floquet engineering, which views light as a time-periodic driving field, has developed as a powerful scheme to directly modulate the Hamiltonian determining the electron/spin states within solid state systems as well as atomic systems [1,2]. Floquet theory claims that when a quantum system is exposed to time-periodic driving, new sets of eigenstates emerge that are separated from each other by integer multiples of $\hbar\Omega$, where $\Omega$ is the driving angular frequency. These are the so-called Floquet-Bloch bands and have been observed for topological insulator surface states [3,4].

Floquet engineering is unique in that it provides a handle to tune the Hamiltonian itself, and therefore offers a way to control the topological properties of quantum systems, as is evident in the concept of "Floquet topological insulators" [5]. One example of this is two-dimensional (2D) honeycomb lattice systems, which are known to host Dirac fermions, under circular driving. Floquet analysis shows that circular driving opens a gap in the otherwise gapless linear dispersion, which is equivalent to the results of the Haldane model [6]. The gap opening is accompanied by the retrieval of a non-zero Chern number (the topological invariant), meaning the system has become topologically nontrivial [7–9]. This concept was first experimentally demonstrated in cold atom systems under circular lattice modulation [10], and soon after in graphene under circularly polarized light (CPL) irradiation [11]. In both cases, the induced topological state was determined through its nonzero Berry curvature (BC), which in the latter case appears in the anomalous Hall effect (AHE). This is because the dc-anomalous Hall conductivity (AHC) is closely related to the BC as [7,8,12]

$$\sigma_{xy} = \frac{e^2}{\hbar} \int \frac{d^d \mathbf{p}}{(2\pi)^d} \sum_n \mathcal{F}_z^n(\mathbf{p}) \times w_n(\mathbf{p}), \tag{1}$$

where $\sigma_{xy}$, $d$, $\mathcal{F}_z^n$, and $w_n$ are the AHC, dimension ($d = 2$ for graphene), BC's $z$-component for the $n$-th band, and the occupancy of the $n$-th band, respectively.

In the case of 2D systems, the lack of band dispersion along the light propagation direction caused one to only see a gap opening at the Dirac point. This is not the case for three-dimensional (3D) Dirac materials, which result in a completely different landscape. The CPL-driven state of 3D Dirac electrons has been extensively studied theoretically, especially under the assumption that the driving photon energy is larger than the system's energy scale (the "high-frequency limit"), due to its simplicity. Within this regime, the CPL induces a chiral gauge field that splits the Dirac cone into two Weyl cones aligned in the light-propagation direction, and the nodes of such cones are called "Floquet Weyl points (FWPs)." [13–15]. The CPL-induced chiral gauge field was recently demonstrated experimentally in the massive 3D Dirac phase of $Co_3Sn_2S_2$ [16]. On the contrary, the effects of resonant driving, i.e., the case where the photon energy lies within the 3D Dirac bands as depicted in Fig.1 (a), is much less investigated despite its unique physics. The emergent topological nodes are not limited to the FWPs, nor are the Chern numbers associated with them limited to $\pm 1$, and different nodes can even show annihilation with each other depending on the driving strength [17,18]. Topological nodes can also be seen in regions far from the Fermi energy, especially where the interband transition is resonant with the driving frequency [19,20]. However, the experimental realization of such topological nodes as well as their effect on quantum geometric responses such as the AHE are highly-nontrivial and unexplored problems.

In this letter, we aim to determine whether or not these topological nodes are achievable in realistic materials/driving schemes and to investigate how they affect the AHE. As a realization of 3D Dirac material, we focus on the elemental bismuth, which is a group V semimetal. The band structure of bismuth based on tight-binding calculations [21] and its Brillouin zone is shown in Fig.1 (b) and (c). The Fermi surface consists of one hole pocket at the $T$-point and three electron pockets at the *L*-points. The electrons are well-described by a massive 3D Dirac Hamiltonian, and its Dirac-like nature is crucial in understanding the colossal diamagnetism and spin Hall effect, two of the many exotic properties the semimetal is known for [22–25].

## II. Method

To measure the AHE induced by CPL-driving in 3D Dirac materials, we prepared a thin film sample of bismuth and conducted mid-infrared (MIR)-pump terahertz (THz) Faraday-probe measurements. A schematic of the experimental setup is shown in Fig.2 (a) (See also appendix A for details). The sample we used was a 30-nm-thick thin film deposited onto a sapphire substrate via radio frequency magnetron sputtering. X-ray diffraction measurements against the sample show that the thin film is polycrystalline with grains oriented along the (003) and (012) directions. The sample was kept in a liquid helium cryostat at 11 K. The pump photon energy was 0.31 eV, which is resonant with the electron-like Dirac bands at the *L*-points but is smaller than the band gap of the hole-like *T*-point. The polarization of the pump pulse was controlled by a liquid crystal retarder so that it is either left or right CPL. The THz Faraday spectroscopy is an efficient probe for the low-frequency responses, which gives us insight into the dc-AHC while also being time-resolved. The polarization of the probe THz pulse is adjusted by a wire grid polarizer before being focused onto the sample (the *x*-axis is defined by this polarization). We measure both the $E_x$- and $E_y$-components of the transmitted THz pulse, and combine the two measurements to obtain the Faraday rotation, which will only be finite under finite AHC. The non-contact nature of this all-optical measurement scheme allows us to measure the CPL-induced AHC while avoiding unintentional symmetry-breaking due to electric contacts. Note that the spot size of the pump pulse (0.23 mm×0.21 mm, 1/*e* radius in terms of the intensity) was comparable to that of the probe pulse (0.24 mm×0.31 mm for 1.0 THz, 1/*e* radius in terms of the electric-field). This results in better pump-probe spatial overlap for higher frequency components of the probe pulse.

## III. Results

In Fig.2 (b), we show the dynamics of transient Faraday rotation ($\theta_F^{peak}$). This quantity is defined as $\theta_F^{peak} = \left(E_{y,L}^{peak} - E_{y,R}^{peak}\right)/\left(E_{x,L}^{peak} + E_{x,R}^{peak}\right)$, where L/R specifies the pump helicity and "peak" represents the fact that it is measured at the peak position of the probe THz pulse. Since the $E_x$-component of the transmitted THz pulse do not show pump-helicity dependence (see Appendix B), $\theta_F^{peak}$ provides a measure of the pump-helicity dependent THz Faraday rotation. The top panel in Fig.2 (b) shows the envelope of the pump pulse intensity (solid line) and the electric field (dashed line), respectively. The full width at half maximum (FWHM) of the CPL-induced Faraday rotation is ~300 fs, which is longer than that of the envelope of the pump pulse intensity (~220 fs) and is comparable to that of the envelope of the electric field (~310 fs). This is due to the sublinear intensity dependence of the CPL-induced AHE, which we will address later. The pump-helicity-dependent Faraday rotation is both finite and instantaneous to the pump pulse, which is a feature expected from Floquet-mediated AHE.

To further understand the instantaneous Faraday rotation, we measured the Faraday rotation spectrum at its peak. The complex Faraday rotation $\widetilde{\Theta}_F(\omega)$ is defined by

$$\widetilde{\Theta}_F(\omega) = \theta_F(\omega) + i\eta_F(\omega) = \frac{\tilde{E}_y(\omega)}{\tilde{E}_x(\omega)}, \quad (2)$$

where $\omega$ is the angular frequency of the probe and $\tilde{E}_x(\omega), \tilde{E}_y(\omega)$ are the Fourier components of the probe THz pulse polarized in the $x$, $y$-direction, respectively. Since we are interested in pump-helicity-dependent Faraday rotation, we take $\tilde{E}_y(\omega) = [\tilde{E}_{y,L}(\omega) - \tilde{E}_{y,R}(\omega)]/2$ and $\tilde{E}_x(\omega) = [\tilde{E}_{x,L}(\omega) + \tilde{E}_{x,R}(\omega)]/2$. $\theta_F(\omega)$ and $\eta_F(\omega)$ are the real and imaginary parts of the complex Faraday rotation, which represent the polarization rotation and ellipticity of the transmitted probe THz pulse, respectively. The results are displayed in Fig.3 (a) and (b).

From the Faraday rotation spectrum, one can further obtain the AHC spectrum $\sigma_{xy}(\omega)$. Assuming that the sample film is optically thin, the two quantities are related to each other by [26]

$$\widetilde{\Theta}_F(\omega) = \frac{Z_0 \sigma_{xy}(\omega) d_{\text{film}}}{1 + n_{\text{subst.}} + Z_0 \sigma_{xx}(\omega) d_{\text{film}}}, \quad (3)$$

where $Z_0, n_{\text{subst.}}, \sigma_{xx}(\omega)$, and $d_{\text{film}}$ is the vacuum impedance, the substrate's refractive index, the longitudinal optical conductivity, and the film thickness, respectively. As shown in Fig.3 (c) and (d), the AHC is dominated by the real part and the imaginary part is almost negligible. The spectra of $\text{Re}[\sigma_{xy}]$ show a slight increase along with the probe photon energy, which is attributed to the improving pump-probe spatial overlap toward higher probe photon energies. Otherwise the spectra of $\text{Re}[\sigma_{xy}]$ is mostly featureless and flat (see Appendix E also). Therefore, we take the average of $\text{Re}[\sigma_{xy}]$ between 2.9 – 9.6 meV as a measure of the CPL-induced AHC.

We plot the AHC as a function of the pump pulse's peak electric field in the sample ($E_{\text{MIR}}$) in Fig.3 (e). Surprisingly, the experimental data show a linear field-strength dependence for the AHC. This cannot be explained by contributions from FWPs, which should yield $\sigma_{xy} \propto E_{\text{MIR}}^2$ (see section IV for further discussion), thus raising the demand to consider the contribution from other topological nodes.

## IV. Discussion
### 1. Effective model for the Floquet states of 3D Dirac electrons

To discuss the role of topological nodes other than the FWPs, we first overview the Floquet band structure under resonant driving, which is shown in Fig.4 (a) and (b). Here the CPL is assumed to propagate along the $z$-axis. This complicated band structure can be understood as replicas of the original Dirac band with the interval of $\hbar\Omega$ overlaid onto each other, along with gap openings where the bands intersect. The colors in Fig.4 (a) and (b) represent the BC's z-component ($\mathcal{F}_z^n$) and the occupancy based on what is called the sudden-approximation ($w_n$, this quantity will be discussed more in detail in the next section) [8], respectively. One can see that there are two types of nodes in the Floquet band structure; the pair of linear dispersions near the Dirac point and the touching near $v_F p_z = \pm \hbar\Omega/2$, i.e., the zone edge of the Floquet bands along the quasi-energy direction. The former nodes are the aforementioned FWPs, which carry the Chern number of unity and act as sources/sinks of BC. While the FWPs are located at BC hotspots, the latter nodes are also located where the BC is concentrated, implying their relevance in the CPL-induced AHC.

To gain a clearer picture of these nodes arising from resonances, we use the following effective Hamiltonian (see appendix C for derivation)

$$H_{\text{eff}}^{\pm}(\mathbf{k}) = -\frac{\Omega}{2} \mp k_z \sigma_z - \frac{1}{A^2+\Omega^2}\begin{pmatrix} \Omega|k|^2 & \pm Ak^2 \\ \pm Ak^{*2} & -\Omega|k|^2 \end{pmatrix}, \quad (4)$$

where $\Omega$ and $A$ are the driving angular frequency and the vector potential's amplitude, respectively ($k = k_x + ik_y$). We chose the unit system $\hbar = e = v_F = 1$. $\pm$ specifies the position of the node ($H_{\text{eff}}^{\pm}$ describes the node near $p_z = \pm \hbar\Omega/2$) and the wavevector $\mathbf{k}$ is measured from it. Note that applying this effective Hamiltonian implicitly neglects the mass and anisotropy of the 3D Dirac bands, which are both present in bismuth [21,27]. The energy dispersion based on $H_{\text{eff}}^{+}$ is displayed in Fig.4 (c) and (d), along with the results of Floquet theory. The band structure obtained from the effective Hamiltonian shows good agreement with that obtained from Floquet theory, showing both linear and quadratic dispersions in the $k_z$-direction and $k_x, k_y$-direction, respectively. This is a feature commonly seen at "double Weyl points," which are topological nodes carrying the Chern number of two. These exotic topological nodes are also discussed in non-driven systems [28–31], though the concept lacks experimental realization. Figures 4 (e) and (f) show the distribution of BC ($\mathcal{F}$) and normalized BC ($\mathcal{F}/|\mathcal{F}|$) around the node at $p_z = \hbar\Omega/2$. We can see that the BC shows a diverging structure around $\mathbf{k} = \mathbf{0}$, indicating the node is topological. By integrating the surface-normal component of the BC over a closed sphere surrounding $\mathbf{k} = \mathbf{0}$, we can obtain the Chern number, which is indeed $\pm 2$. We can therefore say that the nodes due to one-photon resonances can be regarded as double Weyl points, and we shall refer to these nodes hereafter as "Floquet double Weyl points (FDWPs)."

2. Anomalous Hall effect from Floquet double-Weyl points

Having established the topological properties of the FDWPs, we proceed in calculating the AHC stemming from these nodes. For this, one must calculate the BC and occupancy near the FDWPs. The former is obtained straightforwardly [32,33], while the latter requires some assumption, as it depends on the interactions present in the system [34,35]. We assumed the CPL shifts the

Hamiltonian drastically so that the occupancy is determined by the projection from the non-driven state (consistent with Fig.4 (b), see Appendix D also). Under this assumption, we numerically integrate Eq. (1) to obtain the AHC, which is shown by the red line in Fig.4 (g) (see Appendix E for details). Note that the AHC is measured in units of $e^2/\hbar L$, where $L \equiv v_F/\Omega \simeq 7.4 \times 10^{-8}$ cm is the unit length of a unit system where we additionally set $\Omega = 1$. The AHC grows linearly with the field-strength for weaker driving and saturates toward stronger driving. We show the field-strength range corresponding to Fig.3 (e) by the vertical dashed line in Fig.4 (g). In this region, the model suggests a linear field-strength dependence, which is consistent with our experimental observations. The results of model calculations are also plotted along the experimental results as the red line in Fig.3 (e). We can see that the model calculation matches our experimental results not only qualitatively but also quantitatively, disagreeing with only a factor of ~0.22, of which origins we discuss later.

We also compare the contribution from FDWPs with that of FWPs, which are shown by the blue lines in Fig.4 (g) (see Appendix F for details). For the FWPs we use two different assumptions for the occupancy. One is consistent with the assumption made for FDWPs (denoted as SA which stands for "sudden-approximation", dashed line) and the other is that the Fermi-Dirac distribution is formed after the creation of FWPs (denoted as FD which stands for "Fermi-Dirac", solid line). The latter assumption is inconsistent with our calculation for FDWPs, but we have included it nonetheless since it is the common assumption utilized in many theoretical predictions [36,37]. The results for FWPs depend on the Fermi energy which we have set to be 30 meV based on the Fermi energy of bismuth [21]. We plot the same results against $\beta = A^2/\Omega$ in the inset of Fig.4 (g). $\beta$ is the parameter representing the effect of CPL driving at the FWPs [14,15]. The contribution from FWPs is almost negligible for the sudden-approximation while it shows a $\beta$-linear dependence if the Fermi-Dirac distribution is assumed. In either case, contributions from FDWPs are dominant within experimentally achievable field strengths, and neither can explain the linear field-strength dependence observed in experiments.

Strictly speaking, the parameter $\beta$ must exceed the size of the mass term $\Delta$ to form FWPs in the case of massive 3D Dirac bands (see Appendix F also). For $\beta < \Delta$, the doubly-degenerate Dirac bands show splitting without any linear band crossings, while still exhibiting finite BC and thus finite AHE. We note that the strongest driving field achieved in our experiments yields $\beta \simeq 8.9$ meV, which is larger than the mass term of bismuth $\Delta \simeq 7.5$ meV [27]. Therefore, our results indicate that the FDWPs dominate the AHC, even when the Floquet-Weyl semimetal state is achieved.

Finally, we discuss possible factors that may cause the discrepancy between experimental results and model calculations. These include laser-induced heating, carrier scattering [38] (see Appendix E and G also), finite pump pulse width, and spatial pump-probe overlap, which are all likely to reduce the measured AHC. We have also neglected the anisotropy and mass of the Dirac bands, which may lead to corrections to our model calculations. We also note that we have neglected the effect of uneven population formation within the Dirac band induced by the probe THz field, which for graphene happens to be comparable to or exceed that of BC in terms of the AHE [39]. The AHC induced by this mechanism tends to show an intensity-linear power dependence for weak excitation followed by saturation for stronger excitations. The pump-induced carrier density change of bismuth also shows both intensity-linear and saturating behaviors in our experimental range (see Appendix B also), while the CPL-induced AHE shows a linear field-strength dependence across the whole field-strength range investigated in our experiments. We therefore argue that the observed CPL-induced AHE is likely to have an origin distinct from population formation within the Dirac bands. Further research involving microscopic calculations would give a more comprehensive picture of the CPL-induced AHE, which we leave as an open issue for the future.

## V. Conclusion

In conclusion, we conducted MIR-pump THz Faraday-probe measurements of a thin film sample of bismuth, a typical 3D Dirac material. We observed THz Faraday signals that are both helicity-dependent and instantaneous, which are features expected for Floquet-mediated AHE. The power dependence of the AHC along with model calculations suggests that the CPL-induced AHE originates from new topological nodes that emerge when the driving frequency lies within the Dirac bands, which we call FDWPs based on our analysis of their topological properties. Our results not only serve as the first suggestion and the experimental demonstration of FDWPs but can also be regarded as the first realization of any sort of double Weyl state in solid-state systems. We emphasize that FDWPs are not specific to bismuth, but should be a robust and general feature for all 3D Dirac materials whenever they are driven by resonant light. Our results highlight the importance of the frequently overlooked topological aspect of Floquet systems and demonstrate the potential of Floquet systems as platforms to explore exotic topological phenomena, including that of the demonstrated double Weyl state.

**Acknowledgment**

We acknowledge K. Ogawa for assistance in MIR-pump THz Faraday-probe measurements and T. Morimoto for insightful advice on theoretical calculations. We also acknowledge M. Schüler and S. A. Sato for fruitful discussions. This work was supported by JST CREST (Grant No. JPMJCR19T3), Japan. Y. H. was supported by Grant-in-Aid for JSPS Research Fellows (Grant No. 21J20873).

**Appendix A: Experimental details**

Here we give a more detailed explanation of the experimental setup. A schematic of the experimental setup is shown in Fig.5.

We used a Yb: KGW-based regenerative amplifier as our light source (center wavelength: 1030 nm, pulse energy: 2mJ, repetition rate: 3 kHz, pulse duration: 170 fs). The near-infrared (NIR) output pulse was split into three pulses using beam splitters which were used for MIR pulse generation, THz pulse generation, and THz pulse detection, respectively.

The first pulse was converted into a MIR pulse (center wavelength: 4 μm, pulse energy: 6.7 μJ) by an optical parametric amplifier. The MIR pulse went through two ZnSe wire grid polarizers (WGPs) for pulse power adjustment. The polarization of the MIR pulse was controlled by a liquid crystal retarder and the MIR pulse was focused onto the thin film sample by a $CaF_2$ lens. An optical chopper was inserted in the path of the MIR pulse to enable differential measurement, which enhances the signal-to-noise ratio of the pump-induced signal. The pulse duration of the MIR pulse was measured at the sample position by cross-correlation measurements with the NIR pulse using a thin GaSe film. The cross-correlation traces are shown in Fig.2 (b), and by deconvoluting the pulse width of the NIR pulse, one obtains the pulse width of the MIR pulse as 83 fs. The spot size was also measured at the sample position by the knife-edge method, which results were 0.23 mm×0.21 mm (1/*e* radius). The maximum fluence at the sample surface (after the cryostat window) was 1.90 mJ/cm$^2$.

The latter two NIR pulses were used for THz pulse generation and detection. The probe THz was generated by shining the first NIR pulse onto a 380 μm-thick GaP crystal. The generated THz pulse goes through a WGP before being focused onto the sample, which fixes the polarization of the incident THz pulse and defines the *x*-direction necessary for THz Faraday measurements. The transmitted THz pulse goes through two additional WGPs before being focused onto another 380 μm-thick GaP crystal where the electro-optic sampling method was used for THz detection. The WGPs inserted on the detection side have two possible configurations, one for measuring the *x*-component of the transmitted THz pulse and the other for measuring the *y*-component. For the former/latter configuration, the first WGP is set so that it transmits THz pulses that are parallel with/perpendicular to the incident THz polarization. The second WGP is set at a 45°angle between the two configurations of the first WGP. This enables measurements of $E_x$ and $E_y$ that are directly comparable to each other without any additional corrections due to the polarization dependence of the EO-sampling method. The spot size of the THz probe pulse was also measured at the sample position using the knife-edge method, which results were 0.24 mm×0.31 mm, 0.19 mm×0.24 mm, and 0.14 mm×0.20 mm for 1.0 THz, 1.5 THz, and 2.0 THz, respectively (1/*e* radius, electric-field). Note that this is comparable with that of the pump-pulse, which would result in better spatial pump-probe overlap for higher THz probe frequencies.

**Appendix B: Pump-induced change in the longitudinal conductivity**

Here we address the CPL-induced change in the longitudinal conductivity, which appears in the CPL-induced change of the $E_x$-component of the transmitted THz pulse. Note that for the results of this appendix, we have slightly widened the spot size of the pump pulse (0.52 mm×0.53 mm, 1/$e$ radius) to obtain a more meaningful spectrum. The exact results may be slightly affected by this condition change, though the overall behavior of the power dependence should not be altered significantly.

In Fig.6 (a) we display the transient amplitude transmittance measured at the THz peak. For this, we show two different quantities, $\left(\Delta E_{x,L}^{peak} + \Delta E_{x,R}^{peak}\right)/2E_{x,NP}^{peak}$ and $\left(\Delta E_{x,L}^{peak} - \Delta E_{x,R}^{peak}\right)/2E_{x,NP}^{peak}$, where $\Delta E$ is the pump-induced change in the transmitted THz pulse and "NP" (which stands for "no pump") is the transmitted THz electric field when the pump pulse is absent, respectively. The former/latter quantity represents the pump-helicity independent/dependent transmittance change induced by CPL irradiation and is shown in Fig.6 (a) as the red/black line. The helicity-dependent transmittance change is almost negligible since bismuth has no intrinsic time-reversal symmetry breaking. On the contrary, the helicity-independent transmittance change is surprisingly large, approaching almost 30 %.

To understand this drastic change in the THz transmittance, we measured the optical conductivity spectrum at $t_{pp}$ = 2 ps. The real part of the optical conductivity spectrum is shown in Fig.6 (b) by the red dots, along with that obtained without the pump (black circles). For the excited state, the data show a reduction in lower probing energies, which is attributed to the imperfect spatial overlap between the pump and probe. The MIR pumping drastically increases the optical conductivity which indicates efficient carrier population formation. To get a more quantitative picture of this, we fit the higher energy region of the optical conductivity spectrum with the Drude model

$$\sigma_{xx}(\omega) = \frac{\varepsilon_0 \omega_p^2 \tau}{1 - i\omega\tau} \tag{B.1}$$

where $\varepsilon_0, \omega_p, \omega, \tau$ is the vacuum permittivity, the plasma frequency, the probe angular frequency, and the scattering time, respectively. The fitting results are shown along the experimental data by the red and black lines. Since $\omega_p^2$ is proportional to the carrier density we plot the pump-induced change of $\omega_p^2$ in Fig.6 (c) as a function of excitation fluence. The change of $\omega_p^2$ and thus the change in carrier density shows an intensity-linear power dependence for weaker excitation while showing saturation somewhere above 0.2 mJ/cm$^2$. The data shown in Fig.3 (e) represents the AHC both above and below the fluence of 0.2 mJ/cm$^2$ (corresponding to ~0.23 MV/cm) and shows a consistent electric-field linear behavior in both regimes. The fact that the population formation shows both intensity-linear and saturating behaviors while the AHC shows a simple field-linear power dependence indicates that the latter has an origin distinct from population effects.

**Appendix C: Derivation of effective Hamiltonian**

Here we derive the effective Hamiltonian for the FDWPs that we have used for model calculations. We start with the massive 3D Dirac Hamiltonian [14,24]

$$\mathcal{H}_0(\mathbf{p}) = \gamma^0 [\Delta + v_F^x \gamma^x p_x + v_F^y \gamma^y p_y + v_F^z \gamma^z p_z], \tag{C.1}$$

where $\mathbf{p}, \Delta, v_F^{x,y,z}$ and $\gamma^{0,x,y,z}$ are the momentum measured from the Dirac point, the mass term, the Fermi velocity in the $x, y, z$-direction, and the gamma matrices, respectively. Though the electron pockets of bismuth are highly anisotropic, we approximate all the Fermi velocities with their geometric average $v_F = (v_F^x v_F^y v_F^z)^{1/3} \simeq 3.5 \times 10^5$ m/s [40], and adjust our unit system so that $v_F = 1$. For the sake of simplicity, we also adjust our unit system so that $e = \hbar = 1$, which is consistent with the main text. Applying CPL-driving to this system will turn this Hamiltonian time-dependent, which takes the form of

$$\mathcal{H}(\mathbf{p}, t) = \mathcal{H}_0[\mathbf{p} + \mathbf{A}(t)] = \gamma^0 [\Delta + \boldsymbol{\gamma} \cdot \mathbf{p}] + \gamma^0 \boldsymbol{\gamma} \cdot \mathbf{A}(t), \tag{C.2}$$

where $\mathbf{A}(t)$ is the time-dependent vector potential. The effect of CPL-driving is taken into consideration via the Peierls substitution. For CPL with one helicity, the explicit form of the vector potential is given as $\mathbf{A}(t) = (A \cos \Omega t \quad A \sin \Omega t \quad 0)$.

Floquet formalism requires us to calculate the Fourier transform of the time-dependent Hamiltonian, as they are the block matrices with which the Floquet Hamiltonian is constructed. The explicit form of this turns out to be

$$H_{n,m} = \begin{cases} \gamma^0 [\Delta + \boldsymbol{\gamma} \cdot \mathbf{p}] + n\Omega & (n = m) \\ \frac{1}{2} A \gamma^0 (\gamma^x + i\gamma^y) & (n = m + 1) \\ \frac{1}{2} A \gamma^0 (\gamma^x - i\gamma^y) & (n = m - 1) \\ 0 & \text{(otherwise)} \end{cases}. \tag{C.3}$$

The Floquet formalism converts the time-dependent eigenproblem into an infinite-dimension time-independent eigenproblem by using a Hamiltonian that is constructed by an infinite number of such four-by-four matrices. To focus on the one-photon resonant part of the Floquet Hamiltonian, we reduce the Floquet Hamiltonian as follows

$$\mathcal{H}_{\text{reduc.}} = \begin{pmatrix} H_{0,0} & H_{0,-1} \\ H_{-1,0} & H_{-1,-1} \end{pmatrix}. \tag{C.4}$$

This reduced Hamiltonian contains the zero-photon states ($H_{0,0}$), the one-photon emitted states ($H_{-1,-1}$), and the coupling between such states ($H_{0,-1}, H_{-1,0}$). By choosing the Weyl representation this can be explicitly written down as

$$\mathcal{H}_{\text{reduc.}} = \begin{pmatrix} -\mathbf{p}\cdot\boldsymbol{\sigma} & \Delta & -A\sigma_+ & 0 \\ \Delta & \mathbf{p}\cdot\boldsymbol{\sigma} & 0 & A\sigma_+ \\ -A\sigma_- & 0 & -\mathbf{p}\cdot\boldsymbol{\sigma}-\Omega & \Delta \\ 0 & A\sigma_- & \Delta & \mathbf{p}\cdot\boldsymbol{\sigma}-\Omega \end{pmatrix}, \quad (\text{C.5})$$

where $\sigma_\pm = \frac{1}{2}(\sigma_x \pm i\sigma_y)$. If one neglects the mass term ($\Delta \to 0$), this Hamiltonian can be further disentangled into two four-by-four Hamiltonians; one of them being

$$\mathcal{H}^+_{4\times 4} = \begin{pmatrix} p_z & p^* & 0 & A \\ p & -p_z & 0 & 0 \\ 0 & 0 & p_z - \Omega & p^* \\ A & 0 & p & -p_z - \Omega \end{pmatrix}, \quad (\text{C.6})$$

where $p = p_x + ip_y$. The inner portion of $\mathcal{H}^+_{4\times 4}$ implies that there is some band crossing at $p_z = \Omega/2$ (This means we will obtain $H^+_{\text{eff}}$ as our effective Hamiltonian, see main text for its definition). We therefore redefine the wavevector so that is measured from $(p_x \; p_y \; p_z) = (0 \; 0 \; \Omega/2)$. This results in

$$\mathcal{H}^+_{4\times 4} = \begin{pmatrix} k_z + \Omega/2 & k^* & 0 & A \\ k & -k_z - \Omega/2 & 0 & 0 \\ 0 & 0 & k_z - \Omega/2 & k^* \\ A & 0 & k & -k_z - 3\Omega/2 \end{pmatrix}, \quad (\text{C.7})$$

where $\mathbf{k}$ is the redefined wavevector and $k = k_x + ik_y$. For $\mathbf{k} = \mathbf{0}$, the Hamiltonian gives the following eigenenergies and eigenvectors

$$E_{m=1} = -\Omega/2, |m=1\rangle = \begin{pmatrix} 0 & 1 & 0 & 0 \end{pmatrix}^T \quad (\text{C.8})$$

$$E_{m=2} = -\Omega/2, |m=2\rangle = \begin{pmatrix} 0 & 0 & 1 & 0 \end{pmatrix}^T \quad (\text{C.9})$$

$$E_{l=1} = -\Omega/2 + \sqrt{\Omega^2 + A^2}, |l=1\rangle = \begin{pmatrix} \cos\frac{\theta}{2} & 0 & 0 & \sin\frac{\theta}{2} \end{pmatrix}^T \quad (\text{C.10})$$

$$E_{l=2} = -\Omega/2 - \sqrt{\Omega^2 + A^2}, |l=2\rangle = \begin{pmatrix} \sin\frac{\theta}{2} & 0 & 0 & -\cos\frac{\theta}{2} \end{pmatrix}^T. \quad (\text{C.11})$$

The parameter $\theta$ is defined by $\cos\theta = \frac{\Omega}{\sqrt{A^2+\Omega^2}}$. Of the four eigenstates, the ones labeled with $m$ are particularly important since they coincide with that of the FDWP when $\mathbf{k} = \mathbf{0}$. The other two eigenstates labeled with $l$ correspond to the two quasi-energies above and below the FDWP which can be seen in Fig.4 (a)(b).

To construct the effective Hamiltonian, we use the results of perturbation theory, which reads

$$(H_{\text{eff}})_{mn}(x_\mu) = \langle m|\mathcal{H}_{4\times 4}|n\rangle + \langle m|\partial_\mu \mathcal{H}_{4\times 4}|n\rangle x_\mu$$
$$+ \frac{1}{2}\sum_{l=1,2}\left(\frac{\langle m|\partial_\mu \mathcal{H}_{4\times 4}|l\rangle\langle l|\partial_\nu \mathcal{H}_{4\times 4}|n\rangle}{E_m - E_l} + \frac{\langle m|\partial_\nu \mathcal{H}_{4\times 4}|l\rangle\langle l|\partial_\mu \mathcal{H}_{4\times 4}|n\rangle}{E_n - E_l}\right)x_\mu x_\nu, \quad (\text{C.12})$$

where $x_\mu$ is the parameter set used for perturbation expansion. Here, indices $m$ and $n$ are either 1 or 2, which correspond to either one of the two eigenstates we have previously labeled with $m$. The sum in the last term is only conducted for the two eigenstates which we have previously labeled with $l$.

Since we want to expand our Hamiltonian in terms of the wavevector, we choose $x_\mu$ to be $k_z, k$ or $k^*$. Below we list the matrix elements used to carry out this perturbation expansion.

$$\langle m|\mathcal{H}_{4\times4}|n\rangle = \begin{cases} -\Omega/2 & (m=n) \\ 0 & (m\neq n) \end{cases} \tag{C.13}$$

$$\langle m|\partial_{k_z}\mathcal{H}_{4\times4}|n\rangle = \begin{cases} -1 & (m=n=1) \\ 1 & (m=n=2) \\ 0 & (m\neq n) \end{cases} \tag{C.14}$$

$$\langle m|\partial_k \mathcal{H}_{4\times4}|n\rangle = \langle m|\partial_{k^*}\mathcal{H}_{4\times4}|n\rangle = 0 \tag{C.15}$$

$$\langle m|\partial_k \mathcal{H}_{4\times4}|l\rangle = \begin{cases} 0 & (m=1) \\ \sin(\theta/2) & (m=2, l=1) \\ -\cos(\theta/2) & (m=2, l=2) \end{cases} \tag{C.16}$$

$$\langle m|\partial_{k^*}\mathcal{H}_{4\times4}|l\rangle = \begin{cases} 0 & (m=2) \\ \cos(\theta/2) & (m=1, l=1) \\ \sin(\theta/2) & (m=1, l=2) \end{cases} \tag{C.17}$$

$$\langle l|\partial_k \mathcal{H}_{4\times4}|m\rangle = \begin{cases} 0 & (m=2) \\ \cos(\theta/2) & (m=1, l=1) \\ \sin(\theta/2) & (m=1, l=2) \end{cases} \tag{C.18}$$

$$\langle l|\partial_{k^*}\mathcal{H}_{4\times4}|m\rangle = \begin{cases} 0 & (m=1) \\ \sin(\theta/2) & (m=2, l=1) \\ -\cos(\theta/2) & (m=2, l=2) \end{cases} \tag{C.19}$$

By plugging in Eq. (C.13) – (C.19) to Eq. (C.12), one can calculate the effective Hamiltonian which is identical to $H^+_{\text{eff}}$. If we go back to Eq. (C.5) and take the other four-by-four Hamiltonian:

$$\mathcal{H}^-_{4\times4} = \begin{pmatrix} -p_z & -p^* & 0 & -A \\ -p & p_z & 0 & 0 \\ 0 & 0 & -p_z - \Omega & -p^* \\ -A & 0 & -p & p_z - \Omega \end{pmatrix}, \tag{C.20}$$

one can carry out completely parallel calculations to obtain the effective Hamiltonian for the FDWP at $p_z = -\Omega/2$, which results are also shown in the main text.

Since the effective Hamiltonian is in a two-by-two form, it can be expanded in terms of the Pauli matrices. If one expands it as

$$H^\pm_{\text{eff}} = b_0 \sigma_0 + \mathbf{b}^\pm \cdot \boldsymbol{\sigma}, \tag{C.21}$$

the explicit form of $b_0$ and $\mathbf{b}^\pm$ are the following

$$b_0 = -\frac{\Omega}{2}, b^\pm_x = \mp\frac{A(k_x^2 - k_y^2)}{A^2 + \Omega^2}, b^\pm_y = \mp\frac{2Ak_x k_y}{A^2 + \Omega^2}, b^\pm_z = \mp k_z - \frac{\Omega(k_x^2 + k_y^2)}{A^2 + \Omega^2}. \tag{C.22}$$

These will be used in the calculation of BC and occupancy, which is explained in Appendix D.

**Appendix D: Calculation of BC and occupancy**

Here we give results regarding the BC's *z*-component and the occupancy around the FDWPs based on our effective Hamiltonian. Within this two-band model, we shall refer to the band with higher/lower energy as "upper/lower bands."

The BC is [32]

$$\boldsymbol{\mathcal{F}}^{\pm}(\mathbf{k}) = -\frac{1}{2(E^{\pm})^3}\left[b_z^{\pm}(\boldsymbol{\nabla}_{\mathbf{k}}b_x^{\pm} \times \boldsymbol{\nabla}_{\mathbf{k}}b_y^{\pm}) + b_x^{\pm}(\boldsymbol{\nabla}_{\mathbf{k}}b_y^{\pm} \times \boldsymbol{\nabla}_{\mathbf{k}}b_z^{\pm}) + b_y^{\pm}(\boldsymbol{\nabla}_{\mathbf{k}}b_z^{\pm} \times \boldsymbol{\nabla}_{\mathbf{k}}b_x^{\pm})\right] \quad (\text{D.1})$$

where we have defined $E^{\pm} = \sqrt{(b_x^{\pm})^2 + (b_y^{\pm})^2 + (b_z^{\pm})^2}$. This expression gives the BC for the upper band of the two-band model, and that of the lower band is identical but inverted. Plugging in Eq. (C.22) to Eq. (D.1) yields

$$\mathcal{F}_{x,y}^{\pm}(\mathbf{k}) = \mp \frac{k_{x,y}}{(E^{\pm})^3} \frac{A^2(k_x^2 + k_y^2)}{(A^2 + \Omega^2)^2}, \quad \mathcal{F}_z^{\pm}(\mathbf{k}) = \mp \frac{2k_z}{(E^{\pm})^3} \frac{A^2(k_x^2 + k_y^2)}{(A^2 + \Omega^2)^2}. \quad (\text{D.2})$$

The BC distribution is depicted in Fig.4 (e) and (f) for $A = 0.2 \times \hbar\Omega$.

The calculation of the occupancy is much more non-trivial since their details would be strongly affected by the various interactions (electrons, phonons, bath, etc.). Here, we assume that the occupancy of the Floquet band is well represented by the projection from its non-driven counterpart. The physics underlying this assumption is that the strong light pulse causes the Hamiltonian to shift drastically, which is why it is referred to as the "sudden-approximation" in some literature [8]. This assumption also implies that the occupancy distribution is solely determined by the occupancy in the non-driven state, and is not affected by any scattering mechanisms. Analytically, the results of the sudden-approximation can be written down for the FDWP at $p_z = \Omega/2$ as

$$w_{u/l}^{+} = \sum_{i=1,2} \left|\langle \psi_{u/l}^{+} | \psi_{0,i}^{+} \rangle\right|^2 \times w_{0,i}^{+}, \quad (\text{D.3})$$

where $w_{u/l}^{+}, w_{0,i}^{+}, |\psi_{u/l}^{+}\rangle$, and $|\psi_{0,i}^{+}\rangle$ are the occupancy of the upper/lower band in the driven state, the occupancy of the *i*-th band in the non-driven state, the eigenstate of the upper/lower band in the driven state, and the eigenstate of the *i*-th band in the non-driven state. It is worth noting that this formalism of occupancy is time-independent, and should be considered as a time-averaged result rather than a sudden projection from non-driven to driven eigenstates at one certain time. A similar issue was discussed by Dehghani *et al.*, who argued that some scattering mechanism (in their case, electron-phonon scattering) is necessary for such an averaging effect [34]. Without scattering mechanisms the occupancy would become time-dependent, or in more experimental terminology, carrier-envelope phase-dependent. Though our formalism of occupancy seems to contradict itself (because we are using a time-averaged formalism while also assuming various scattering mechanisms are negligible), we argue that since the carrier-envelope phase of our pump pulse is random, the time-averaging can occur during data accumulation regardless of what happens within the sample.

The eigenstate for the driven state $|\psi^+_{u/l}\rangle$ can be analytically written down as

$$|\psi^+_{u/l}\rangle = \frac{1}{\sqrt{2E^+(E^+ \mp b_z^+)}}\begin{pmatrix} b_x^+ - ib_y^+ \\ \pm E^+ - b_z^+ \end{pmatrix}. \tag{D.4}$$

For eigenstates of the non-driven state, we chose $|\psi^+_{0,i=1}\rangle = (1 \quad 0)^{\mathrm{T}}$ and $|\psi^+_{0,i=2}\rangle = (0 \quad 1)^{\mathrm{T}}$. These are not energy eigenstates but are a natural choice of eigenstates nonetheless since the former represents the non-driven valence band and the latter represents the non-driven conduction band shifted by the energy of $\hbar\Omega$. The associated occupancies of these eigenstates are $w^+_{0,i=1} = 1$ and $w^+_{0,i=2} = 0$, since we are working in a regime where the Fermi energy and thermal broadening are sufficiently smaller than the photon energy. Plugging this in with Eq. (D.3) gives us

$$w^+_{u/l} = \frac{1}{2}\left(1 \pm \frac{b_z^+}{E^+}\right). \tag{D.5}$$

An identical expression holds for the FDWP located near $p_z = -\Omega/2$.

**Appendix E: Details on the numerical calculation of the AHC**

Here we explain details regarding the numerical calculations of the AHC. As explained in the main text, we calculate the AHC based on a formula similar to Eq. (1). For our numerical calculations, we additionally set our unit system so that $\Omega = 1$. We start by calculating the contribution of one of the FDWPs, which is described by $H^+_{\mathrm{eff}}$. We denote this by $\sigma^+_{xy}$, which can be written down as

$$\sigma^+_{xy} = \frac{e^2}{\hbar L}\int \frac{d^3\mathbf{k}}{(2\pi)^3} \mathcal{F}^+_z (w^+_u - w^+_l), \tag{E.1}$$

where $L \equiv v_{\mathrm{F}}/\Omega \simeq 7.4 \times 10^{-8}$ cm is the unit length defined by our choice of unit system. The other FDWP gives an identical contribution. Therefore, the total AHC turns out to be $\sigma_{xy} = 3 \times 2 \times \sigma^+_{xy}$, where factors three and two account for the number of electron pockets and the number of FDWP per electron pocket, respectively. We numerically integrate the above expression within $-1/2 \leq k_x, k_y, k_z \leq 1/2$, where the effective model reasonably reproduces the Floquet bands. Since $E^+ = 0$ at $\mathbf{k} = \mathbf{0}$, the integrand diverges at the FDWP. We assume the integrand is zero at $\mathbf{k} = \mathbf{0}$ to avoid the whole expression being ill-defined.

We can also extend this calculation to finite frequencies and include effects of finite scattering as well based on the Kubo formula,

$$\sigma^+_{xy}(\omega) = \frac{e^2}{\hbar L}\int \frac{d^3\mathbf{k}}{(2\pi)^3} \frac{w^+_u - w^+_l}{2E^+} \times \left[\frac{\langle\psi^+_u|v_x|\psi^+_l\rangle\langle\psi^+_l|v_y|\psi^-_u\rangle}{-2E^+ + \omega + i\gamma} + \frac{\langle\psi^+_l|v_x|\psi^+_u\rangle\langle\psi^+_u|v_y|\psi^-_l\rangle}{2E^+ + \omega + i\gamma}\right] \tag{E.2}$$

where $\omega$ and $\gamma$ are the probing angular frequency and the scattering rate, respectively. $v_{x,y}$ is the velocity operator in the $x$, $y$-direction, which is defined as $v_{x,y} \equiv \partial H_{\mathrm{eff}}/\partial k_{x,y}$. Explicit calculation using Eq. (C.22), (D.2), and (D.4) shows that Eq. (E.2) can further be simplified as

$$\sigma^+_{xy}(\omega) = \frac{e^2}{\hbar L}\int \frac{d^3\mathbf{k}}{(2\pi)^3} \frac{(2E)^2}{(2E)^2 - (\omega + i\gamma)^2} \mathcal{F}^+_z(w^+_u - w^+_l). \tag{E.3}$$

The FDWP at $p_z = -\Omega/2$ also gives an identical contribution. Note that Eq. (E.3) indeed reduces to Eq. (E.1) if one takes the limit of $\omega, \gamma \to 0$. We again numerically integrate Eq. (E.3) within $-1/2 \leq k_x, k_y, k_z \leq 1/2$ to obtain $\sigma_{xy}^+(\omega)$, and multiply it by six to obtain the total AHC.

The results are shown in Fig. 7 for various scattering rates and driving field strengths. The Drude model fit of the equilibrium optical conductivity yields $\gamma \simeq 29$ THz (see Appendix G also) which corresponds to ~0.06 $\Omega$, but we note that the scattering rate of carriers in the Floquet bands may deviate from this value. The spectral range of the probe (0.7 – 2.3 THz, or 2.9 – 9.6 meV) corresponds to $0.009 \leq \omega \leq 0.031$, which is indicated by the black vertical dashed lines. Within the measurable spectral range, all the AHC spectra show common characteristics such as the real part being featureless and the imaginary part being smaller compared to the real part, regardless of the scattering rate or the field strength. These features are consistent with the experimental results shown in Figs.3 (c) and 3 (d).

We note that the scattering rate can affect the AHC even in the dc-limit ($\omega \to 0$). This is not because the finite scattering makes the Floquet formalism less valid [38], but a consequence that is purely derivable from the Kubo formula. Reduction of AHC due to finite scattering may also be one of the factors causing the experimental results to appear smaller than our model calculations. For Fig.4 (g) we show the results of calculations based on Eq. (E.1) which assumes $\omega, \gamma \to 0$ to examine the upper bound value of the light-induced AHC. Using the results of Eq. (E.1) as a measure of the CPL-induced AHE enables us to compare contributions from FDWPs and FWPs without additional assumptions of carrier scattering rates, which we discuss in Appendix F.

**Appendix F: Comparison between contributions from FWPs and FDWPs**

In the main text, we have discussed that the observed CPL-induced AHE has an FDWP origin. Here we also calculate the contributions from FWPs and compare them with the results shown in Fig.4 (e).

We start by reviewing the derivation of the effective Hamiltonian that describes the FWPs [13–15]. The derivation is based on the assumption that the driving frequency is sufficiently larger than the energy scale we are focusing on. This neglects the photon absorbed/emitted states of the Floquet Hamiltonian, and the only effect the Floquet formalism gives is the perturbation of the non-driven Hamiltonian via the off-diagonal blocks of the Floquet Hamiltonian. The effective Hamiltonian thus reads

$$H_{\text{eff}}(\mathbf{p}) = \mathcal{H}_0(\mathbf{p}) + \frac{1}{\Omega}[H_{-1,0}, H_{0,-1}] = \mathcal{H}_0(\mathbf{p}) - \frac{A^2}{\Omega}\gamma^0\gamma^z\gamma_5 \equiv \mathcal{H}_0(\mathbf{p}) - \beta\gamma^0\gamma^z\gamma_5. \quad (\text{F}.1)$$

The notations are consistent with the main text and previous appendices. The CPL-induced term is equivalent to that representing the application of a chiral gauge field in the $p_z$-direction. By choosing the Weyl representation, the four-by-four matrix form of this Hamiltonian is given as:

$$H_{\text{eff}}(\mathbf{p}) = \begin{pmatrix} -p_z - \beta & -p^* & \Delta & 0 \\ -p & p_z + \beta & 0 & \Delta \\ \Delta & 0 & p_z - \beta & p^* \\ 0 & \Delta & p & -p_z + \beta \end{pmatrix}. \quad (F.2)$$

The resulting quasi-energies from this Hamiltonian are $\pm\varepsilon_\pm(\mathbf{p})$, where

$$\varepsilon_\pm(\mathbf{p}) = \sqrt{p_x^2 + p_y^2 + \left(\sqrt{p_z^2 + \Delta^2} \pm \beta\right)^2}. \quad (F.3)$$

The CPL-induced term makes the Dirac bands split, and if $\beta > \Delta$, FWPs will emerge at $p_z = \pm\sqrt{\beta^2 - \Delta^2}$. Strictly speaking, FWPs only appear under this condition, and otherwise we would have split massive Dirac bands that all possess finite BC. We use the term FWP rather loosely in the main text and refer to contributions from these bands as "FWP contributions" as well since they have the same physical origin.

We proceed in calculating the AHC arising from this effective Hamiltonian. Instead of conducting analytical calculations, we apply the Fukui-Suzuki-Hatsugai method [41] and calculate the BC numerically. The occupancy is calculated based on two different assumptions. One is that the Fermi-Dirac distribution is formed with some Fermi energy after the Floquet bands are formed, and the other is that the occupancy is determined by the projection from the non-driven state ("sudden-approximation"). The latter is consistent with model calculations regarding the FDWPs. The former contradicts the assumption we have made for FDWP model calculations, but we have included it here as a reference. We note that it may be justifiable to use different assumptions for different portions of the Floquet spectrum since the interactions that driven electrons would be subject to may vary, but further research is necessary for the exact determination of the nonequilibrium distribution.

The AHC arising from the FWPs is calculated based on Eq. (1). We numerically integrate the expression in the range of $-\Omega/2 \leq p_x, p_y, p_z \leq \Omega/2$ after setting $\Omega = 1$. We have limited the calculations for FWPs within the range of $0 \leq A \leq 1/2$, since going beyond this range would result in the annihilation of FWPs [17], making the four-band model irrelevant. As additional parameters, we added the mass term $\Delta \simeq 7.5$ meV, the Fermi energy $E_F \simeq 30$ meV [21,27], and the temperature of the initial Fermi-Dirac distribution ~20 K. We note that the AHC from FDWPs does not require any assumption of the Fermi energy since our photon energy is much larger than the Fermi energy. In units of $\hbar\Omega$, the mass term is $\Delta \simeq 0.024$, and the corresponding threshold for FWP generation would be $\beta = 0.024$ or $A = 0.16$. We notice no discontinuity across this threshold in terms of the AHC and therefore refer to contributions from the four-band model as "FWP contributions," regardless of whether or not $\beta > \Delta$ holds.

**Appendix G: Effect of carrier scattering**

Here we address the effect of carrier scattering within our thin film sample. In Fig.8 we show the optical conductivity spectrum measured at 25 K. The relatively flat spectrum indicates that the scattering rate of carriers within bismuth is higher than the spectral range available for our THz detection setup (0.5 – 2.5 THz, or 2.1 – 10.3 meV). The accurate determination of the scattering rate is difficult in this situation, but we fit the optical conductivity spectrum by the Drude model (Eq. (B.1)), which results are plotted along with the experimental data. The extracted scattering time is $\tau \simeq 34$ fs, which corresponds to the scattering rate of $\gamma = \tau^{-1} \simeq 29$ THz. Given that the driving frequency of the MIR pulse was 75 THz, we can expect the carriers to be driven a few cycles per one scattering event. Therefore, the scattering is expected to reduce Floquet-originated signals, but not suppress them entirely.

Figures

Fig. 1

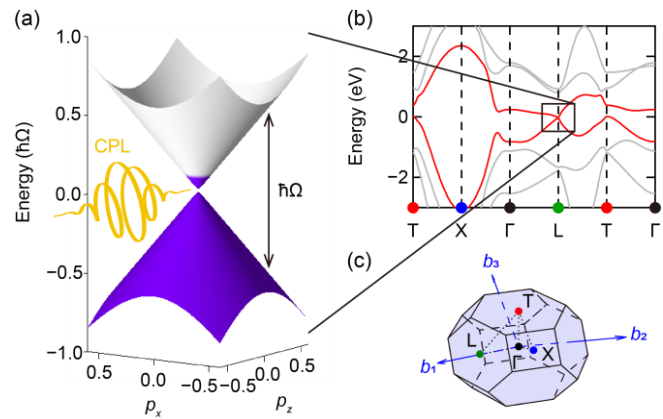

(a) Schematic of massive 3D Dirac bands. Occupied states are colored in purple, which can also be seen in the conduction band due to finite Fermi energy. CPL with the photon energy that is within the 3D Dirac bands is applied to this system. (b), (c) Band structure near the Fermi level and Brillouin zone of elemental bismuth. The bands forming the electron/hole pockets are highlighted in red. The Dirac bands are located in the electron-like $L$-points.

Fig. 2

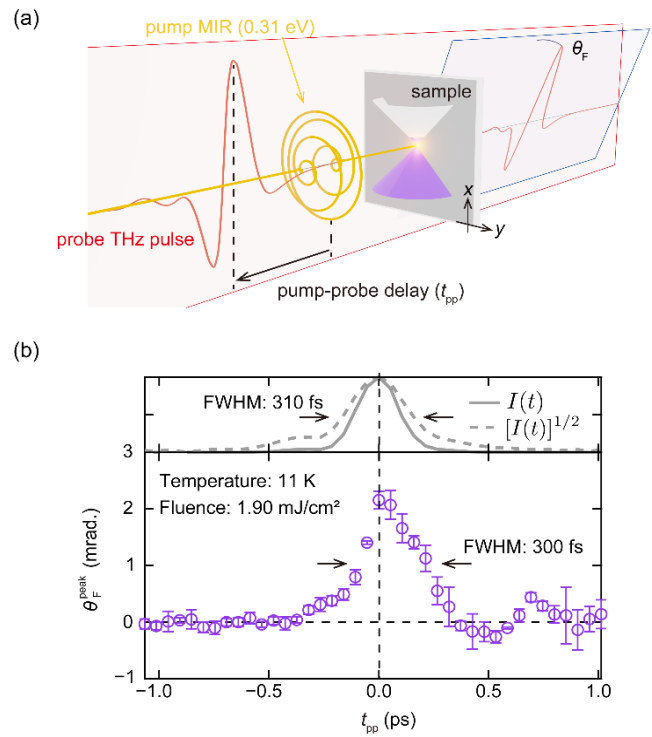

(a) Schematic of the experimental setup. A circularly polarized MIR pulse (orange) and a linearly (*x*-) polarized THz pulse (red) are irradiated onto the bismuth thin film sample. The arrival time difference between the two pulses defines the pump-probe delay time $t_{pp}$. (b) Dynamics of the transient Faraday rotation. The solid/dashed gray traces are the envelope of the pump pulse in terms of intensity $I(t)$/electric field $[I(t)]^{1/2}$, respectively.

Fig. 3

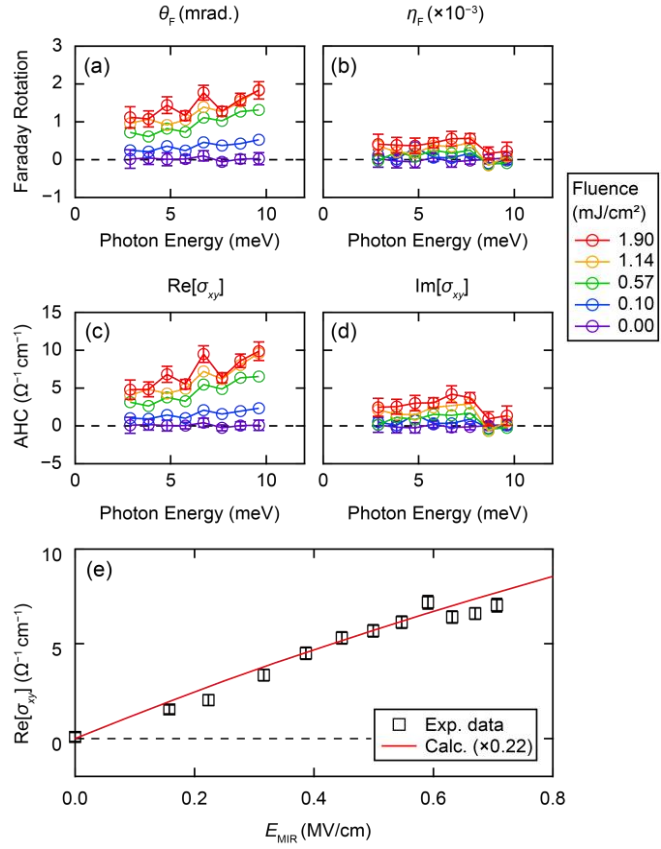

(a), (b) Real and imaginary parts of the Faraday rotation spectrum measured at $t_{pp}$ = 0 ps. (c), (d) Real and imaginary parts of the AHC measured at $t_{pp}$ = 0 ps. The colors represent various pump fluences. (e) Power dependence of the CPL-induced AHE. The squares are the average of the real part of the AHC spectra in the range of 2.6 – 9.6 meV. The red line is the result of model calculations scaled by the factor of 0.22. See the main text for details of the model calculation.

Fig. 4

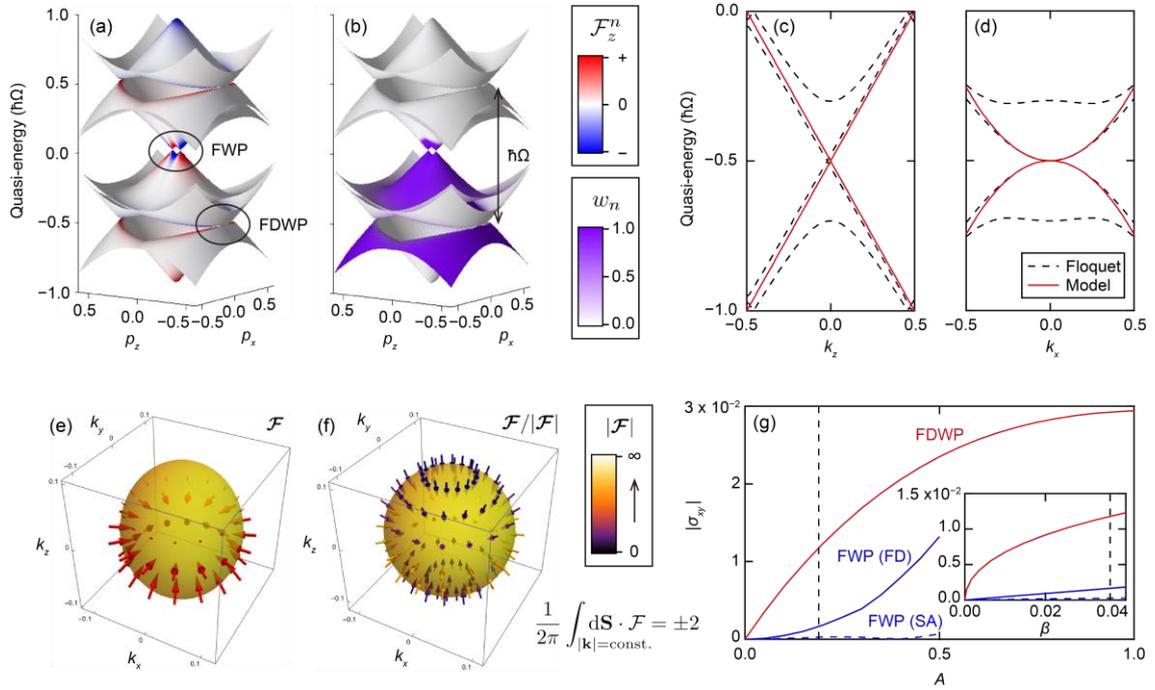

(a), (b) Floquet bands of the 3D Dirac bands. The color scales in (a) and (b) represent BC's z-component and occupancy, respectively. Two different types of nodes are present (FWPs and FDWPs), both located in BC hotspots. (c), (d) Band structure obtained from the effective Hamiltonian (red line). The results of Floquet theory (black dashed lines) are shown for comparison. (e), (f) BC distribution around the FDWP at $p_z = \hbar\Omega/2$. The size of the arrows in (e) represents the intensity of BC, and its normalized version is shown in (f). The FDWP acts as sources/sinks of BC, carrying the Chern number of two. (g) Field-strength dependence of the absolute value of AHC based on numerical calculations. The AHC is measured in units of $e^2/\hbar L$ and the field strength $A$ is measured in units of $\Omega$. The red line shows the results for the FDWPs while the blue lines show the results for the FWPs. The black dashed line shows the field-strength region corresponding to that shown in Fig.3 (e). The inset shows the same results plotted against $\beta = A^2/\Omega$.

Fig. 5

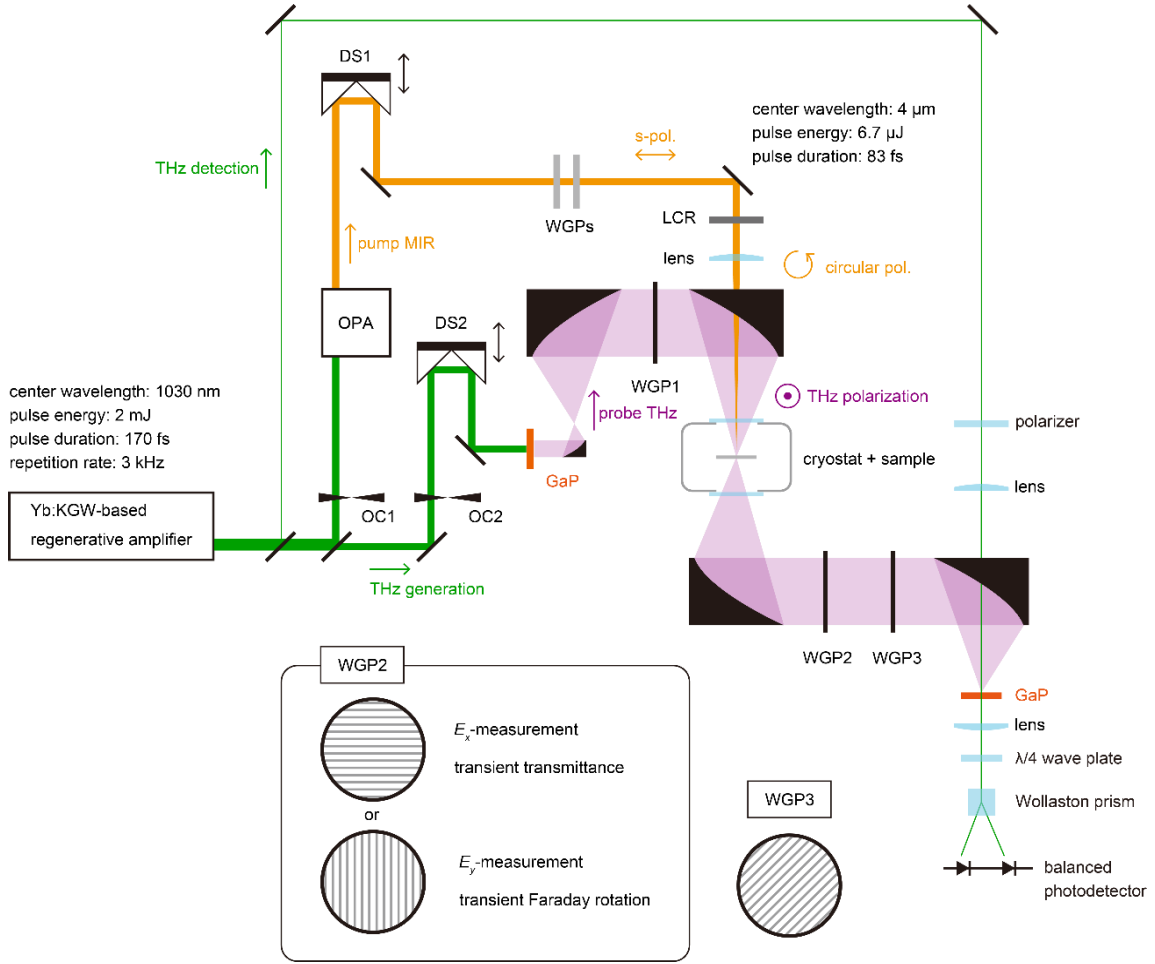

Fig.5. Schematic of the experimental setup. OC: optical chopper, DS: delay stage, WGP: wire grid polarizer, LCR: liquid crystal retarder.

Fig. 6

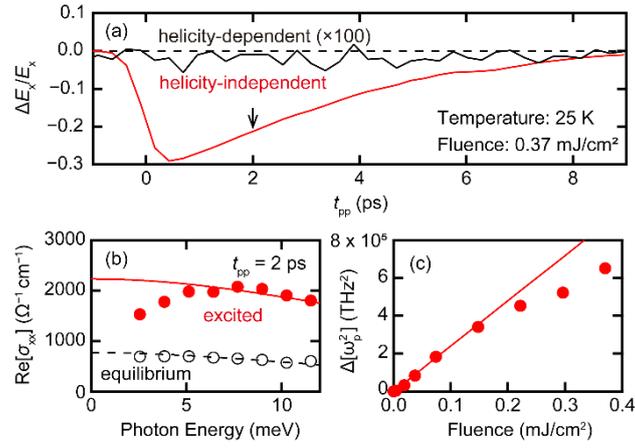

(a) Dynamics of the amplitude transmittance induced by the MIR pump. The definition of $t_{pp}$ is consistent with the main text. The pump-helicity-dependent component is negligible while the pump-helicity-independent component shows a drastic decrease. See Appendix B for definitions of the two quantities. (b) Real part of the optical conductivity w/ (red) and w/o (black) MIR pumping. The symbols are the experimental data while the lines are the results of fitting based on the Drude model. The low energy region of the excited state shows reduction due to imperfect spatial overlap between the pump and probe. (c) Pump-induced change of $\omega_p^2$, where $\omega_p$ is the plasma frequency. The red line shows the result of a linear fit of the data from weaker regions.

Fig. 7

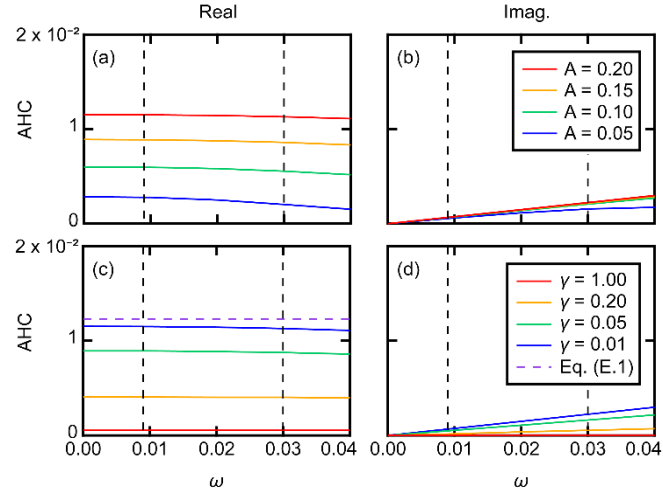

Real and imaginary parts of the AHC spectrum based on the effective Hamiltonian for various field strengths ((a), (b)) and scattering rates ((c), (d)). The probe angular frequency ($\omega$), the field strength ($A$), and the scattering rate ($\gamma$) are all measured in units of $\Omega$ and the AHC is measured in units of $e^2/\hbar L$. The black dashed lines indicate the spectral range corresponding to that of Figs.3 (c) and 3 (d). The purple dashed line in (c) indicates the AHC obtained by taking the limit of $\omega, \gamma \rightarrow 0$, i.e., that obtained based on Eq. (E.1).

Fig. 8

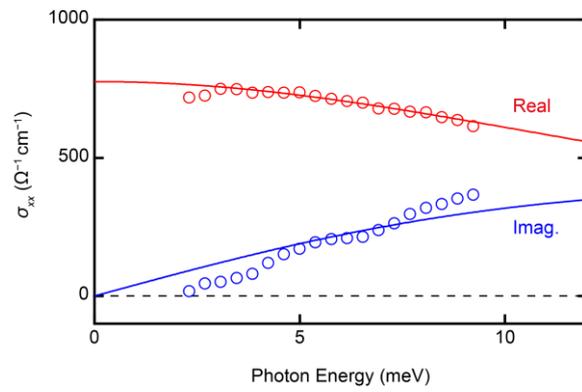

The real and imaginary parts (red and blue, respectively) of the equilibrium optical conductivity of bismuth. The symbols show the experimental data and the traces show the results of fitting based on the Drude model.